\documentclass[11pt,a4paper]{article}
\pdfoutput=1
\usepackage{jcappub}

\title{Low Variance at large scales of WMAP 9 year data}
\author[a,b]{A. Gruppuso,}
\author[c,d,e,a]{P. Natoli,}
\author[f]{F. Paci,}
\author[a,b]{F. Finelli,}
\author[g,a]{D. Molinari,}
\author[a]{A.~De~Rosa}
\author[a,c,h]{and N. Mandolesi}

\affiliation[a]{INAF-IASF Bologna, Via P. Gobetti 101, I-40129, Bologna, Italy}
\affiliation[b]{INFN, Sezione di Bologna, Via Irnerio 46, I-40126 Bologna, Italy}
\affiliation[c]{Dipartimento di Fisica e Scienze della Terra, Universit\`a degli Studi di Ferrara, \\ via Saragat 1, I-44100 Ferrara, Italy}
\affiliation[d]{INFN, Sezione di Ferrara, via Saragat 1, I-44100 Ferrara, Italy}
\affiliation[e]{Agenzia Spaziale Italiana Science Data Center, \\ c/o ESRIN, via Galileo Galilei, Frascati, Italy}
\affiliation[f]{SISSA - Scuola Internazionale Superiore di Studi Avanzati, \\ via Bonomea 265, I-34136 Trieste, Italy}
\affiliation[g]{Dipartimento di Fisica e Astronomia, Universit\`a degli Studi di Bologna, \\ viale Berti Pichat 6/2, I-40127 Bologna, Italy}
\affiliation[h]{Agenzia Spaziale Italiana, viale Liegi 26, I-00198 Roma, Italy}
\emailAdd{gruppuso@iasfbo.inaf.it}
\emailAdd{natoli@fe.infn.it}
\emailAdd{fpaci@sissa.it}
\emailAdd{finelli@iasfbo.inaf.it}
\emailAdd{molinari@iasfbo.inaf.it}
\emailAdd{derosa@iasfbo.inaf.it}
\emailAdd{mandolesi@iasfbo.inaf.it}
\abstract{We use an optimal estimator to study the variance of the WMAP 9 CMB field at low resolution, in both temperature and polarization.  Employing realistic Monte Carlo simulation, we find statistically significant deviations from the $\Lambda$CDM model in several sky cuts for the temperature field. For the considered masks in this analysis, which cover at least the $54\%$ of the sky, the WMAP 9 CMB sky and $\Lambda$CDM are incompatible at $ \ge 99.94\%$ C.L. at large angles ($>5^\circ$). We find instead no anomaly in polarization. As a byproduct of our analysis, we present new, optimal estimates of the WMAP 9 CMB angular power spectra from the WMAP 9 year data at low resolution. }
\arxivnumber{1304.5493}
\begin{document}
\maketitle

\section{Introduction}
\label{intro}
The cosmic microwave background (CMB) anisotropy field can be used to probe cosmology to high precision, as shown by the  Wilkinson Microwave Anisotropy Probe (WMAP) 9 years results \cite{Hinshaw:2012fq} and by the very recent {\sc Planck} cosmological results (see \cite{Ade:2013xsa} and references therein).  
CMB data have given a significant contribution in setting up the $\Lambda$ cold dark matter ($\Lambda$CDM) cosmological concordance model. 
The latter establishes a set of basic quantities for which CMB observations and other cosmological and astrophysical data-sets 
agree\footnote{See, however, \citep{Ade:2013lmv} for a tension concerning the  $\Omega_m$ estimate  from {\sc Planck} CMB and galaxy clusters data.}: spatial curvature close to zero;  $\sim 68.5 \%$ of the cosmic density in the form of Dark Energy; $\sim 26.5\%$ in cold dark matter; $\sim 5\%$ in baryonic matter; and non perfectly scale invariant adiabatic, primordial perturbations compatible with Gaussianity \citep{Ade:2013lta,Ade:2013ydc}.
The largest angular scales of the CMB anisotropies have remained largely primordial, having evolved only in amplitude since the end of the inflation. They can be used to probe the physics of the very early Universe. 
Accurate estimates of their amplitude is crucial to extract information about such primordial phases \citep{Ade:2013uln}. 

In this paper we provide new, temperature and polarization, estimates of the CMB Angular Power Spectra (APS) from the WMAP 9 year low resolution data
using a Quadratic Maximum Likelihood (QML) method \citep{Tegmark:2001zv}. 
We have used {\it BolPol}, an efficient, fully polarized, implementation of the QML method. {\it BolPol}  has been already applied 
to WMAP 5  \citep{Gruppuso:2009ab} and WMAP 7  \citep{Gruppuso:2010nd} low resolution data. 
At large angular scales, QML estimates are superior to the pseudo spectrum techniques normally used in the CMB world  
\citep{master,saha,polenta,grain} because of their smaller intrinsic variance \citep{Molinari2013}. 
The temperature WMAP 9 analysis has been derived using a $C^{-1}$ method which extends to high multipole \citep{Bennett:2012fp}.
{\it BolPol} is fully optimal, but computational considerations limit its application to smaller multipoles. 
This is fully adequate for the low resolution analysis presented here.
For other applications of our QML implementation see \citep{Paci:2010wp,Paci:2013gs}, where it is used to constrain the CMB hemispherical power asymmetry in WMAP 5 and 7 data respectively, and \cite{Gruppuso:2011ci}, where the low resolution spectrum of the cosmological birefringence angle has been derived 
for the first time. 
For an application to {\sc Planck} data see \citep{Planck:2013kta}.
See also \cite{Schiavon:2012fc}, where a similar implementation of our QML code has been applied to detect the cross-correlation between CMB temperature anisotropies and large scale structure surveys.

In this paper, we provide new APS estimates from WMAP 9 year low resolution data. At the same time, we check for their stability against Galactic masking. 
Specifically, we look at the variance of the temperature and polarization maps. The variance of the CMB temperature field has been studied in previous works (see \cite{Cruz:2010ud} for WMAP 5 and 7 year data and 
\cite{Monteserin:2007fv} for WMAP 3 year data). 
Here we focus  on the lowest multipoles  of WMAP 9 year data, we adopt a new estimator  (based on {\it BolPol})  and consider also polarization. 

The paper is organized as follows.
In Section \ref{descriptionscode} we briefly describe the QML method and our  implementation,  {\it BolPol}.
Section \ref{dataset} is devoted to the description of the WMAP 9 year low resolution data set under consideration. 
APS estimates of this data set are given in Section \ref{estimates} where the consistency of all the spectra with what expected in the $\Lambda$CDM model is evaluated
through a $\chi^2$ analysis. In the same Section the stability of the obtained estimates is studied against the Galactic masking.
The analysis of the variance is presented in Section \ref{lowvariance}.
Our conclusions are drawn in Section \ref{conclusions} where some speculations about a possible origin for this effect are set forth.
A proof of concept of a Bayesian approach is given in Appendix \ref{HZ}.


\section{Angular power spectra estimator}
\label{descriptionscode}

In order to evaluate the APS we adopt the QML estimator,
introduced in \citep{Tegmark:1996qt}  and extended to polarization in \citep{Tegmark:2001zv}. 
In this section we describe the essence of such a method. 

Given a map in temperature and polarization ${\bf x=(T,Q,U)}$, the QML provides estimates
$\hat {C}_\ell^X$ - with $X$ being one of $TT, EE, TE, BB,
TB, EB$ - of the APS as: 
\begin{equation}
\hat{C}_\ell^X = \sum_{\ell' \,, X'} (F^{-1})^{X \, X'}_{\ell\ell'} \left[ {\bf x}^t
{\bf E}^{\ell'}_{X'} {\bf x}-tr({\bf N}{\bf
E}^{\ell'}_{X'}) \right]
\, ,
\end{equation}
where the $F_{X X'}^{\ell \ell '}$ is the Fisher matrix, defined as
\begin{equation}
\label{eq:fisher}
F^{\ell\ell'}_{X X'}=\frac{1}{2}tr\Big[{\bf C}^{-1}\frac{\partial
{\bf C}}{\partial
  C_\ell^X}{\bf C}^{-1}\frac{\partial {\bf C}}{\partial
C_{\ell'}^{X'}}\Big] \,,
\end{equation}
and the ${\bf E}^{\ell}_X$ matrix is given by
\begin{equation}
\label{eq:Elle}
{\bf E}^\ell_X=\frac{1}{2}{\bf C}^{-1}\frac{\partial {\bf C}}{\partial
  C_\ell^X}{\bf C}^{-1} \, ,
\end{equation}
with ${\bf C} ={\bf S}(C_{\ell}^X)+{\bf N}$ being the global covariance matrix (signal plus noise contribution). 

Although an initial assumption for a fiducial power spectrum $C_{\ell}^X$ is needed, the QML method provides unbiased estimates of the power spectrum contained 
in the map regardless of the initial guess,
\begin{equation}
\langle\hat{C}_\ell^X\rangle= \bar C_\ell^{X} \,,
\label{unbiased}
\end{equation}
where the average is taken over the ensemble of realizations (or, in a practical test, over Monte Carlo 
realizations ex-tracted from $\bar C_\ell^{X}$).
On the other hand, the covariance matrix associated to the estimates,
\begin{equation}
\langle\Delta\hat{C}_\ell^X
\Delta\hat{C}_{\ell'}^{X'} \rangle= ( F^{-1})^{X \, X'}_{\ell\ell'} \,,
\label{minimum}
\end{equation}
does depend on the initial assumption for $C_\ell^X$: 
the closer the guess to the true power spectrum is, the closer are the error bars to minimum variance.
According to the Cramer-Rao inequality, which sets a limit to the accuracy of an estimator, Eq. (\ref{minimum}) tells us that 
the QML has the smallest error bars.  The QML is then an `optimal' estimator.

This method has been implemented in a F90 code, named {\it BolPol}.
Further details can be found in \citep{Gruppuso:2009ab}.

\section{Data set}
\label{dataset}
In this Section we describe the WMAP 9 year data set that we have considered.
We use the temperature ILC map smoothed at $9.1285$ degrees and reconstructed at HealPix\footnote{http://healpix.jpl.nasa.gov/}
 \citep{gorski} resolution $N_{side}=16$, 
the foreground cleaned low resolution maps and the noise covariance matrix in $(Q,U)$ publicly available at the LAMBDA website\footnote{http://lambda.gsfc.nasa.gov/} 
for the frequency channels Ka, Q and W as considered by \cite{Bennett:2012fp} and \cite{Larson:2010gs} for the low $\ell$ analysis. 
These frequency channels have been co-added as follows \citep{Jarosik:2006ib,Gruppuso:2010nd}
\begin{equation}
m_{tot} = C_{tot} (C_{Ka}^{-1} m_{Ka} + C_Q^{-1} m_Q+ C_V^{-1} m_V)
\, ,
\end{equation}
where $m_{i}$, $C_i$ are the polarization maps  and covariances (for $i=Ka, Q$ and $V$) and
\begin{equation}
C_{tot}^{-1} = C_{Ka}^{-1} + C_{Q}^{-1} +C_{V}^{-1} 
\, .
\end{equation}
This polarization data set has been extended to temperature considering the ILC map.
We have added to the temperature map a random noise realization with variance of $1 \mu K^2$ as suggested in \citep{Dunkley:2008ie}.
Consistently, the noise covariance matrix for TT is taken to be diagonal with variance equal to $1 \mu K^2$.

\section{Estimates}
\label{estimates}

\begin{figure}
\centering
\includegraphics[width=135mm]{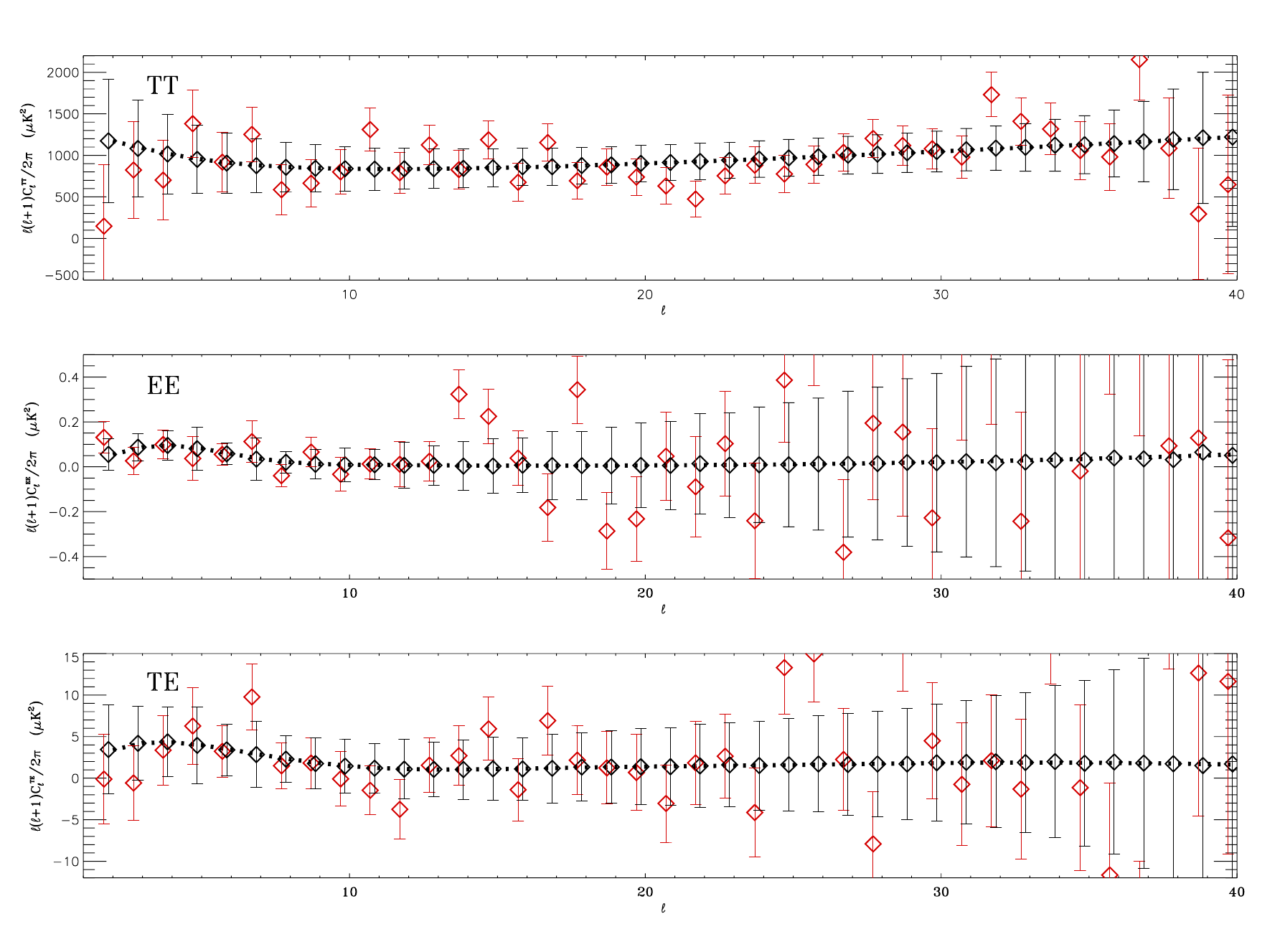}
\includegraphics[width=135mm]{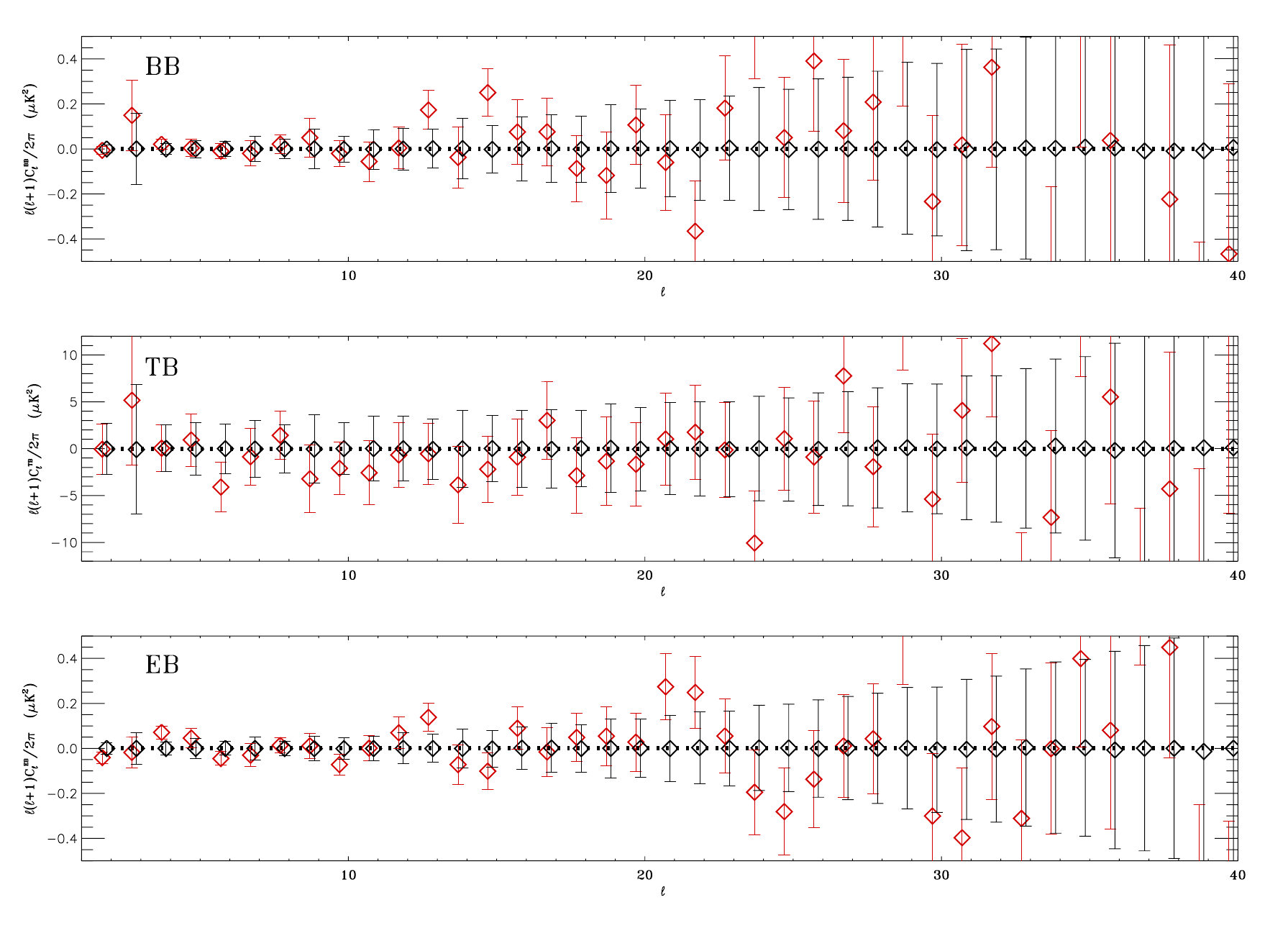}
\caption{WMAP 9 year CMB spectra at low resolution. ILC in Temperature, masked with kq85 mask and the co-addition of Ka, Q and V channel in Polarization masked with the P06 mask.
From top to bottom, TT, EE, TE, BB, TB and EB spectrum. For each spectrum we show $\ell (\ell +1) C_{\ell}/ 2 \pi$ (given in $\mu$K$^2$) versus the multipole order $\ell$.
Red estimates are for the WMAP 9 year maps. Black estimates are the averages of $10000$ Monte-Carlo simulations where the CMB is randomly extracted from
a $\Lambda$CDM model (shown as dotted lines) and the noise is obtained through a Cholesky decomposition of the noise covariance matrix.
See also the text.}
\label{uno}
\end{figure}

Taking into account the kq85 mask for the temperature map\footnote{Best-fit monopole and dipole have been subtracted from the observed ILC map through the HealPix routine {\sc remove-dipole} \cite{gorski}.} 
and the P06 mask for polarization maps we estimate all the 6 CMB spectra of WMAP 9 year data with {\it BolPol}, see red symbols of Fig.~\ref{uno}.
In the same Figure we also show the averages of $10000$ ``CMB plus noise'' Monte-Carlo (MC) simulations (see black symbols) where the CMB is randomly extracted from
a $\Lambda$CDM model, drawn as dotted line, and the noise is obtained through a Cholesky decomposition of the noise covariance matrix.
Consistently with WMAP data, also the temperature simulated maps have been smoothed at that scale, i.e. $9.1285$ degrees.

\subsection{ $\chi^2$ consistency with $\Lambda$CDM model}
\label{chi2section}

\begin{figure}
\centering
\includegraphics[width=61mm]{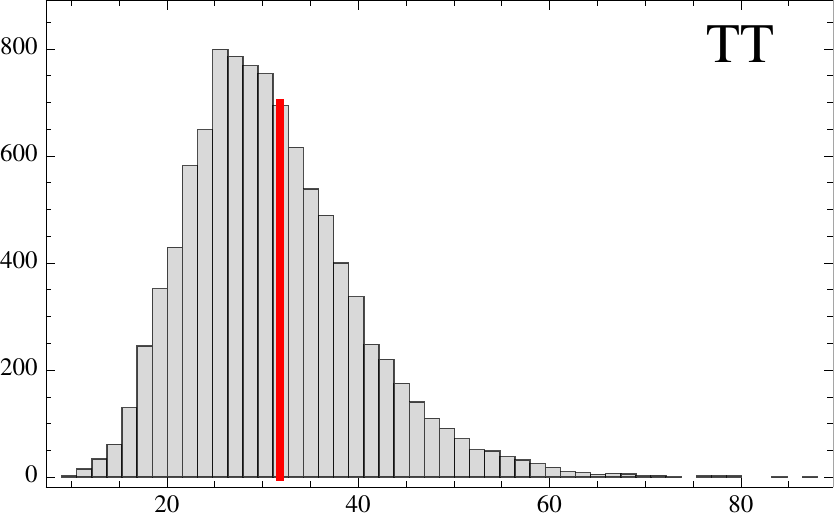}
\includegraphics[width=61mm]{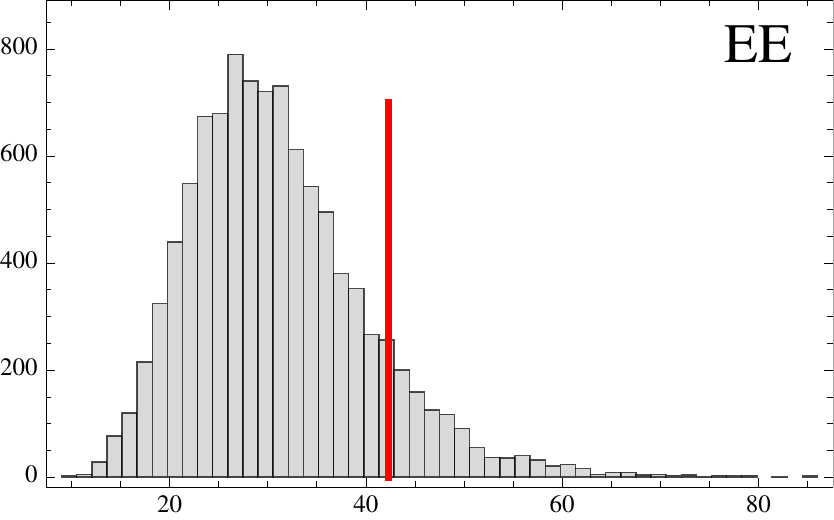}

\includegraphics[width=61mm]{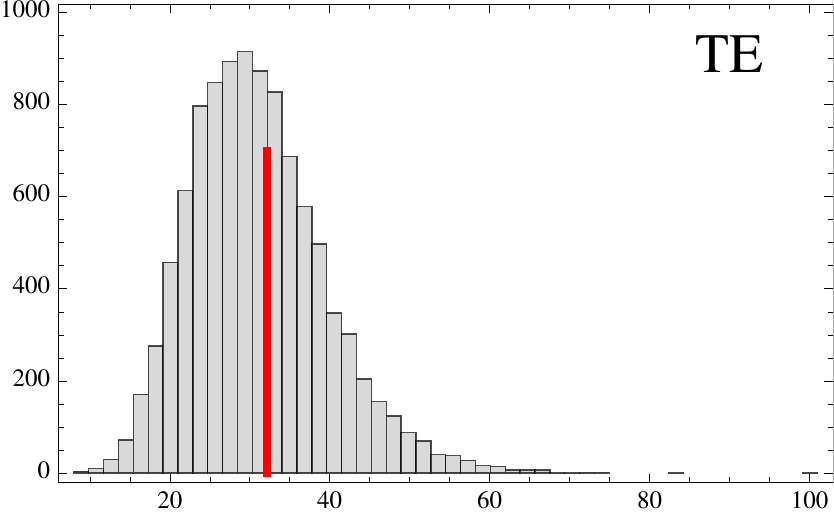}
\includegraphics[width=61mm]{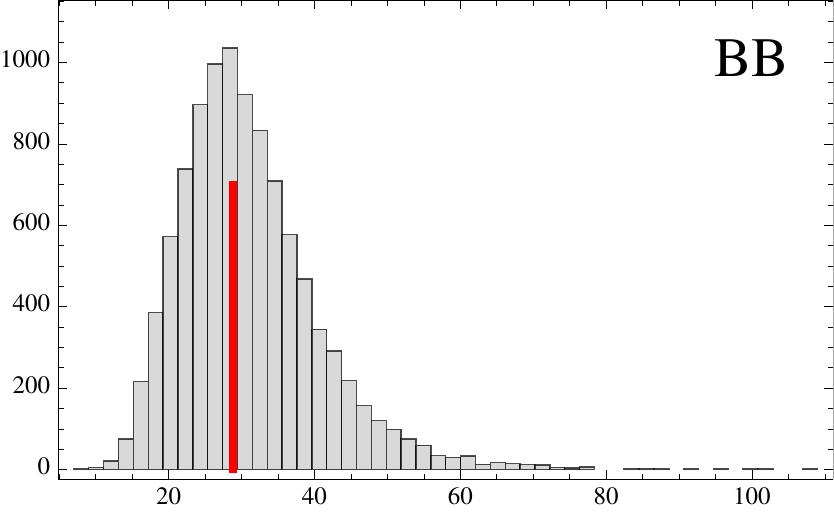}

\includegraphics[width=61mm]{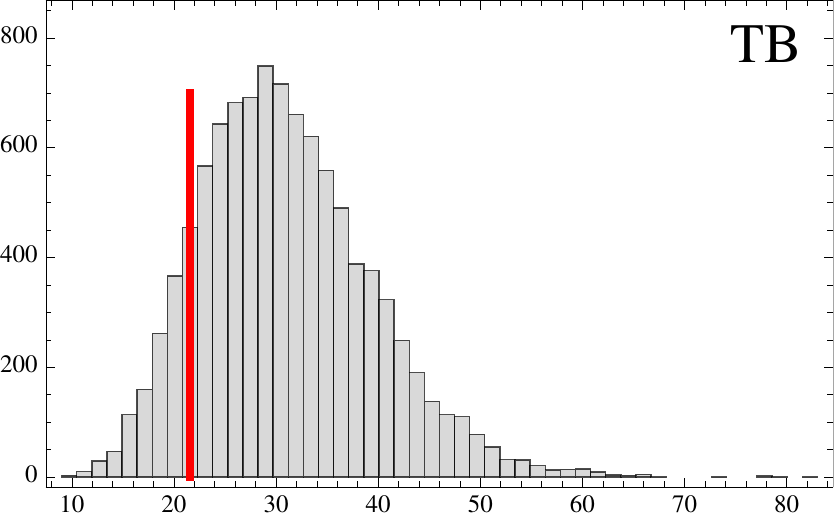}
\includegraphics[width=61mm]{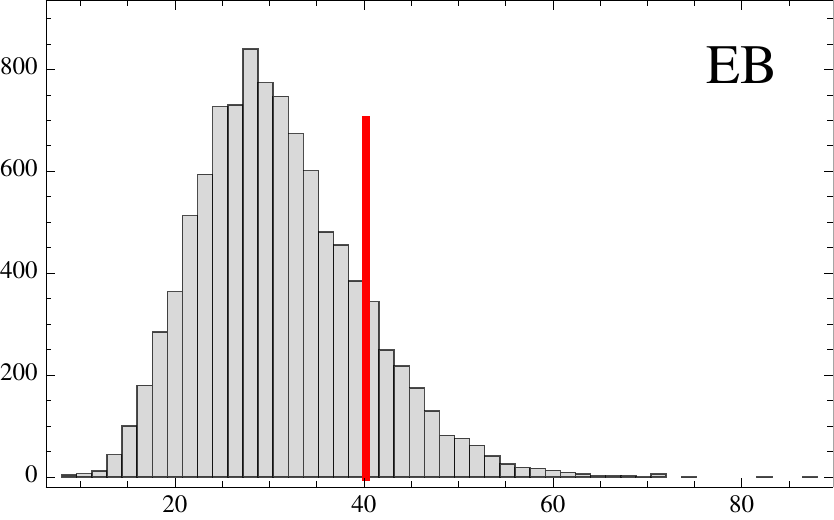}

\caption{$\chi^2$ test for consistency with $\Lambda$CDM model. Each panel shows the counts versus $\chi^2$.
The histograms are built through Monte Carlo simulations. The red lines are the $\chi^2$ values we obtain for the WMAP 9 year low resolution data.
See also the text.}
\label{due}
\end{figure}

In order to evaluate the consistency with the $\Lambda$CDM model (identified in the following with a set of $C_{\ell}^{th}$ for each spectrum) we adopt a frequentist approach,
see Fig.~\ref{due}. 
For each spectrum 
(TT, EE, TE, BB, TB and EB) we plot the histogram
of the $\chi^2_i$ values defined as
\begin{equation}
\chi^2_{i} = \sum_{\ell,\ell^{\prime}=2,32}(C_{\ell,i}-C_{\ell}^{th}) M_{\ell \ell^{\prime}}^{-1} (C_{\ell^{\prime},i}-C_{\ell^{\prime}}^{th})
\, ,
\label{chi2}
\end{equation}
where $i=1,10000$ runs over the MC realizations and where the covariance matrix $M_{\ell \ell^{\prime}}$ is defined as 
\begin{equation}
M_{\ell \ell^{\prime}}=\langle (C_{\ell,i}-C_{\ell}^{th})(C_{\ell^{\prime},i}-C_{\ell^{\prime}}^{th})\rangle
\, ,
\label{defM}
\end{equation}
with $\langle \cdot \cdot \cdot \rangle$ standing for the ensamble average.
These histograms will be compared with the $\chi^2_{\rm WMAP9}$ obtained with the WMAP 9 year data for each spectrum, defined as
\begin{equation}
\chi^2_{\rm WMAP9} = \sum_{\ell, \ell^{\prime}=2,32}(C_{\ell,{\rm W9}}-C_{\ell}^{th}) M_{\ell \ell^{\prime}}^{-1} (C_{\ell^{\prime},{\rm W9}}-C_{\ell^{\prime}}^{th})
\, ,
\label{chi2WMAP9}
\end{equation}
where $C_{\ell,{\rm W9}}$ are the WMAP 9 year spectra.
Fig.~\ref{due} shows that the WMAP 9 year low resolution data appear to be well consistent with the $\Lambda$CDM model defined by
the WMAP 9 fiducial model, represented by the dotted line shown in Fig.~\ref{uno}.
See Table \ref{chi2table} for a quantitative analysis.

\begin{table}
\centering
\caption{Percentage to obtain a $\chi^2$ value from a random extraction of $\Lambda$CDM model smaller than what observed by WMAP 9.}
\label{chi2table}
\begin{tabular}{cc}
\hline
Spectrum & Percentage \\
\hline
TT & 59.26  \\
EE & 88.99  \\
TE & 59.25  \\
BB & 46.13  \\
TB & 12.22  \\
EB & 85.73  \\
\hline
\end{tabular}
\end{table}

\subsection{Stability of the results versus Galactic masking}
\label{maskanalysis}

\begin{figure}
\includegraphics[width=80mm]{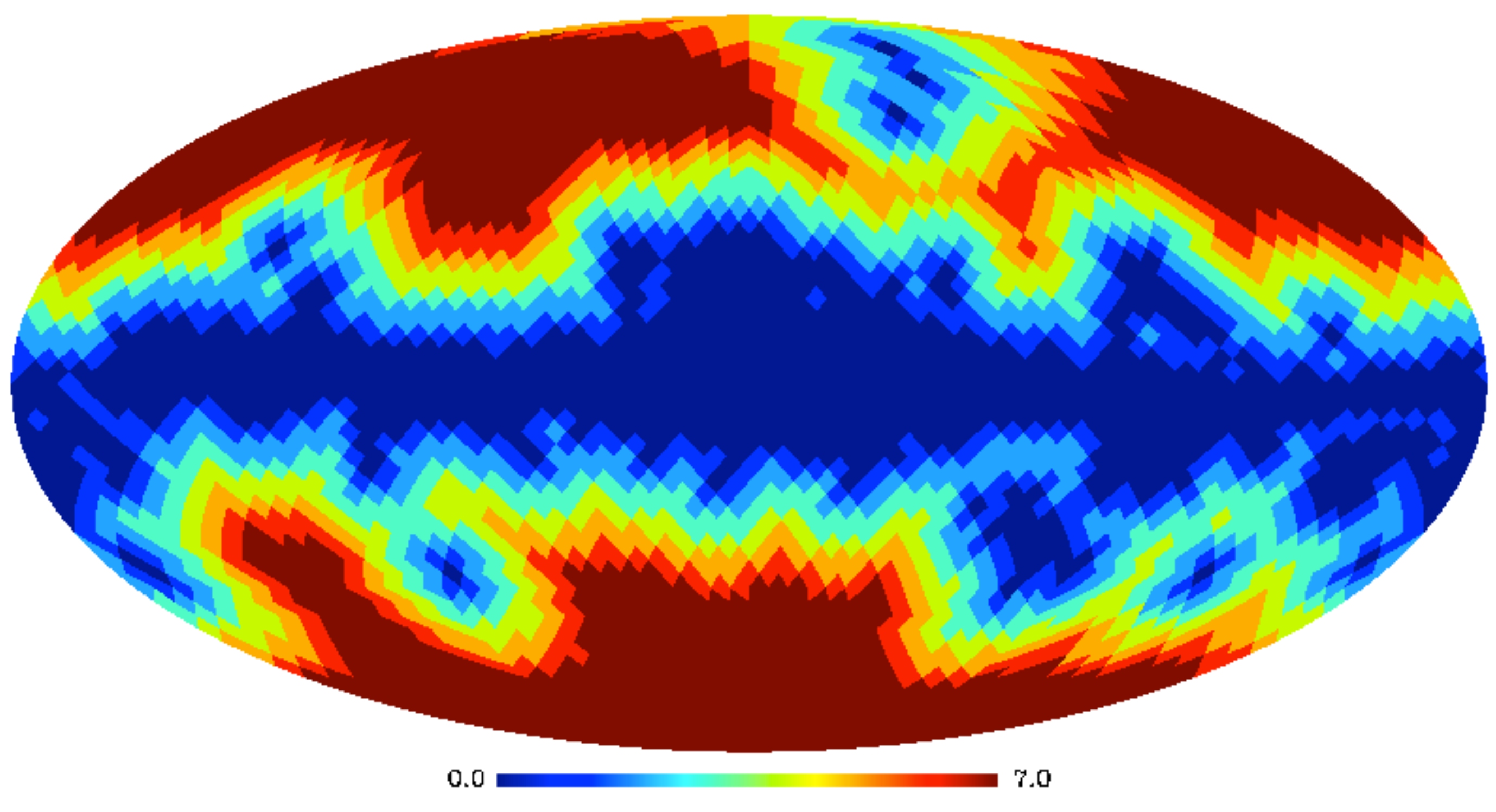}
\includegraphics[width=80mm]{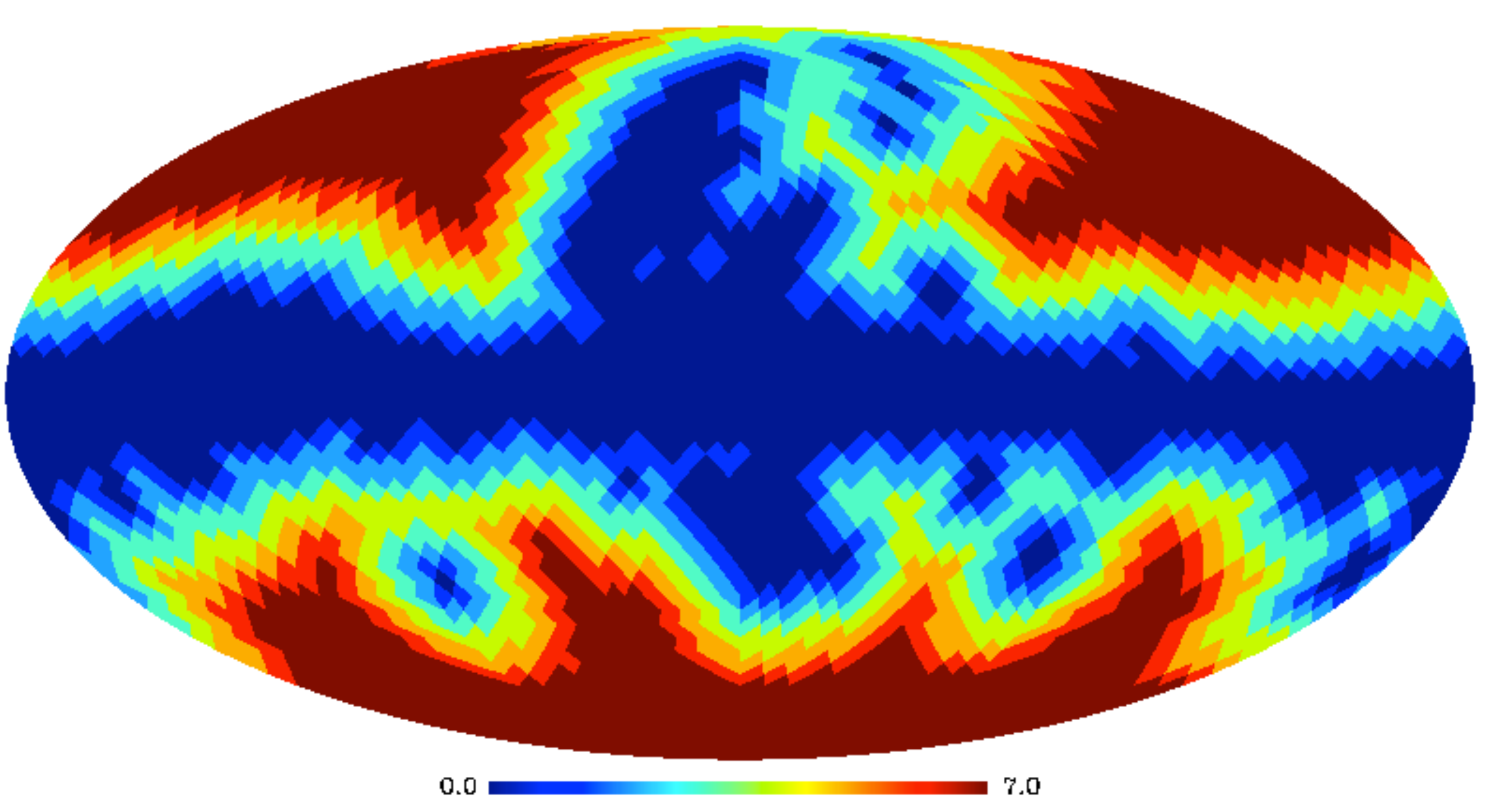}
\caption{Galactic masks. Left panel: Masks we adopt in Temperature. Right panel: Masks we adopt in Polarization.
Case ``a'' is masked where the value 0 is. Case ``b'' is masked where the values 1 and 0 are. Case ``c'' is masked where the values 2, 1 and 0 are.
Case ``d'' is masked where the values 3, 2, 1 and 0 are. Case ``e'' is masked where the values 4, 3, 2, 1 and 0 are. Case ``f'' is masked where the values 5, 4, 3, 2, 1 and 0 are.
Case ``g'' is masked where the values 6, 5, 4, 3, 2, 1 and 0 are. 
See also Table \ref{maskstabel}.}
\label{tre}
\end{figure}

In this Subsection we test the stability of the WMAP 9 year data estimates 
considering various 
Galactic masks, see Fig.~\ref{tre}.
More specifically, we extend the edges of the kq85 temperature mask and the P06 polarization mask by 4, 8, 12, 16, 20 and 24 degrees.
In Table \ref{maskstabel} we provide the sky fraction for each considered case and for reference we rename with the letters ``a'', ``b'', ``c'', ``d'', ``e'', ``f'' and ``g'' the seven
cases we take into account. Case ``a'' is what is already considered in Subsection \ref{chi2section}.

\begin{table}
\centering
\caption{Observed sky fraction of the considered cases. See also Fig.~\ref{tre}.}
\label{maskstabel}
\begin{tabular}{cccc}
\hline
Case & Extension wrt & Observed sky & Observed sky \\
 & kq85 and P06  [deg] & fraction in T & fraction in P \\
\hline
a & +0 & 0.78  & 0.73 \\
b & +4 & 0.68 & 0.65 \\
c & +8 & 0.56  & 0.53 \\
d & +12 & 0.46 & 0.43 \\
e & +16 & 0.36 & 0.34 \\
f & +20 & 0.28 & 0.27 \\
g & +24 & 0.22 & 0.20 \\
\hline
\end{tabular}
\end{table}

\begin{figure*}
\centering
\includegraphics[width=130mm]{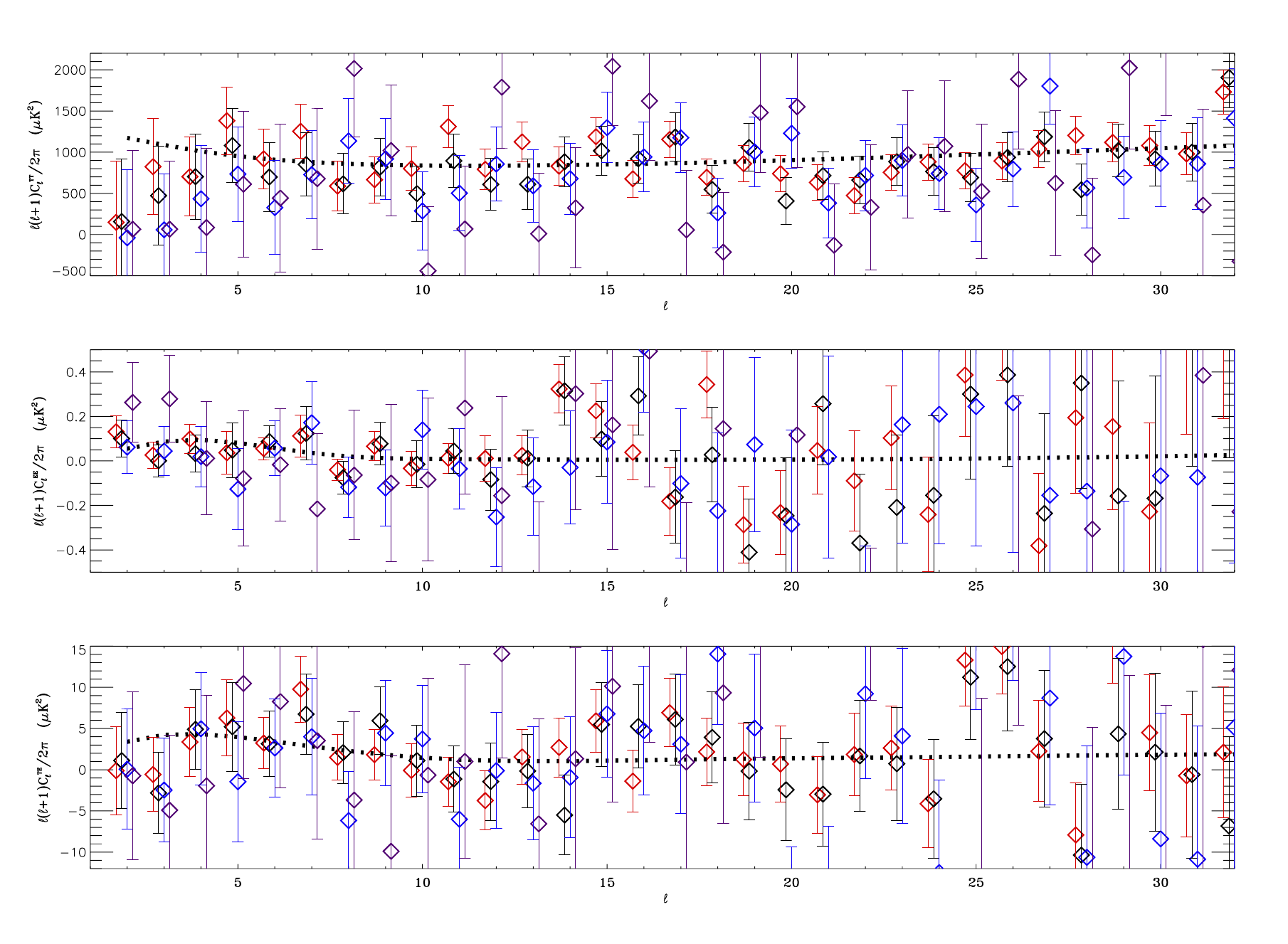}
\includegraphics[width=130mm]{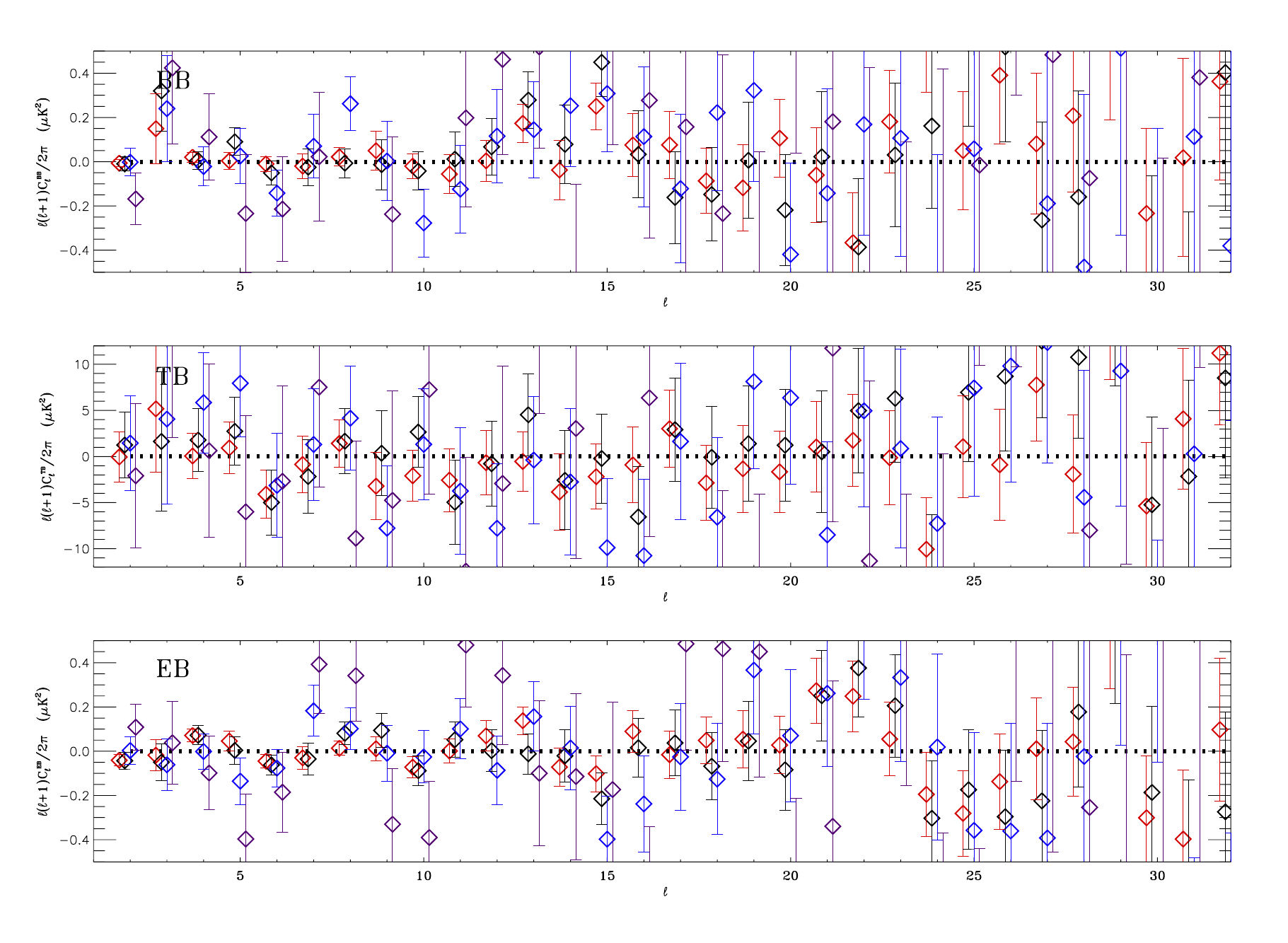}
\caption{WMAP 9 year CMB spectra at low resolution. Red symbols for case ``a'', Black symbols for case ``c'', Blue symbols for case ``e'' and Violet symbols for case ``g''.
See also Table \ref{maskstabel}. From top to bottom, TT, EE, TE, BB, TB and EB spectrum. For each spectrum we show $\ell (\ell +1) C_{\ell}/ 2 \pi$ (given in $\mu$K$^2$) versus the multipole order $\ell$.
See also the text.}
\label{quattro}
\end{figure*}

\begin{figure}
\centering
\includegraphics[width=95mm]{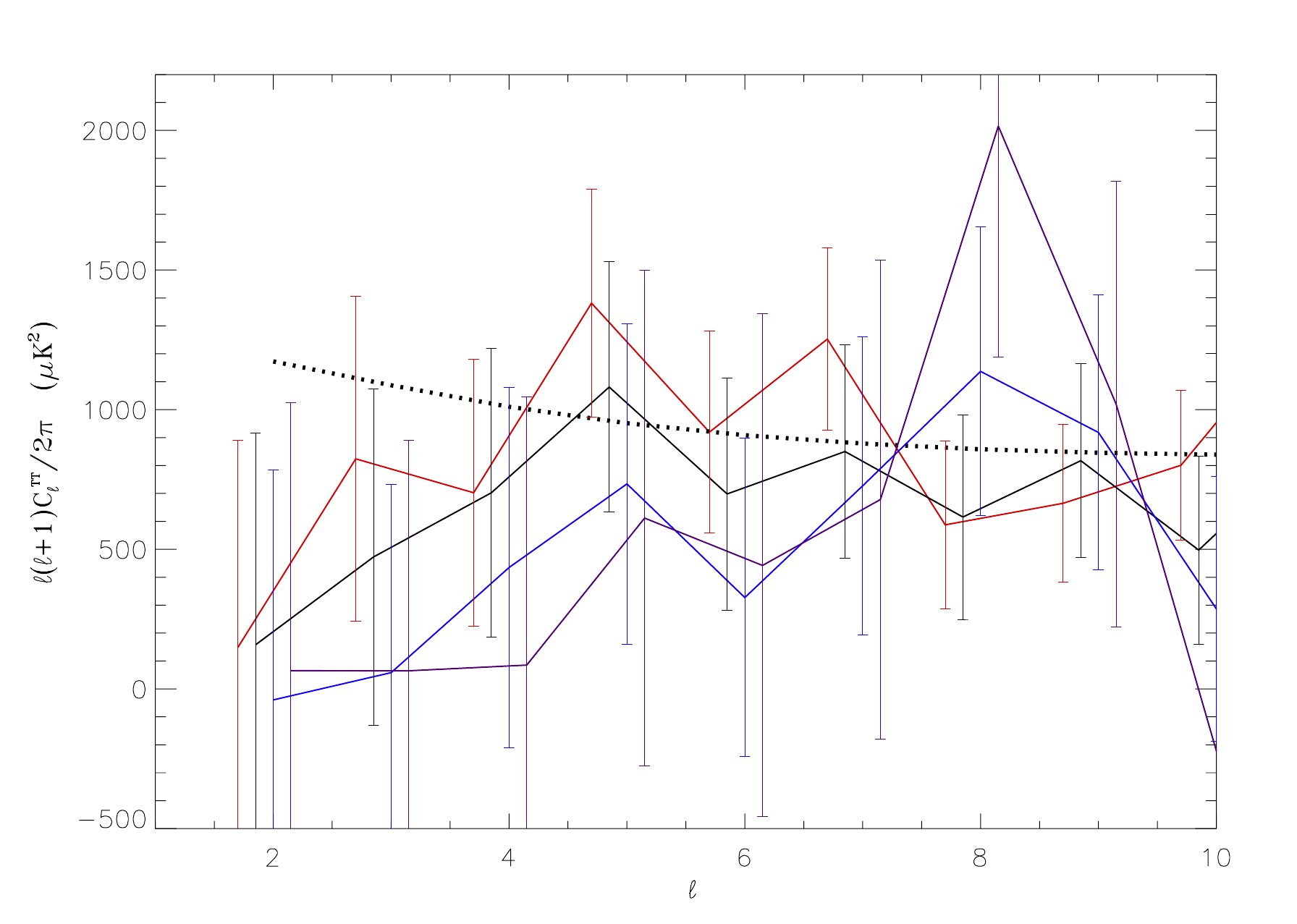}
\caption{Zoom of the TT spectrum given in Fig.~\ref{quattro} at the lowest multipoles. See also caption of Fig.~\ref{quattro}.}
\label{cinque}
\end{figure}

In Fig.~\ref{quattro} we show the six spectra for four out of seven cases, namely case ``a'', ``c'', ``e'' and ``g''
\footnote{For sake of clarity in Fig.~\ref{quattro} we do not show all the cases but just four out of seven.}.
The overall consistency is good. As expected the more extreme case, i.e. ``g'' case, shows the larger fluctuations. 
However, the lowest multipoles of the TT spectrum show a pretty clear trend: increasing the temperature mask, we obtain a lowering  of the power at the largest scales from $\ell=2$
to $\ell \sim 7$. Fig. ~\ref{cinque} shows a zoom on the TT panel of Fig.~\ref{quattro} for the lowest multipoles, where the 
estimates are now joined with broken lines in order to make this behavior clearer. 

\section{Variance analysis}
\label{lowvariance}

In order to measure such a deviation from WMAP 9 fiducial model we do not take into account the estimator given in Eq.~(\ref{chi2})
since it is not sensitive to a lack of power because of its quadratic expression in terms of distance from the theoretical model $C_{\ell}^{th}$.
Therefore we decide to naturally consider the expressions for the Variance of the CMB fluctuations in pixel space, which can be written as follows, see e.g. Appendix A of \citep{Tegmark:2001zv}
\begin{equation}
\langle \delta T ^2 \rangle = \sum_{\ell \ge 2}^{\ell_{max}} \left( {{2 \ell +1} \over {4 \pi}} \right) C_{\ell}^{TT} \, ,
\label{VARTT} 
\end{equation}
\begin{equation}
\langle Q ^2 \rangle = \sum_{\ell \ge 2}^{\ell_{max}} \left( {{2 \ell +1} \over {8 \pi}} \right) \left( C_{\ell}^{EE} + C_{\ell}^{BB} \right) \, ,
\label{VARQQ}
\end{equation}
\begin{equation}
\langle U ^2 \rangle = \langle Q ^2 \rangle \, 
\label{VARUU}
\end{equation}
where $\delta T$, is the fluctuation of the temperature map and $Q$ and $U$ are the Stokes parameters maps.

\subsection{Temperature analysis}
\label{temperatureanalysissection}

\begin{figure}
\centering
\includegraphics[width=45mm]{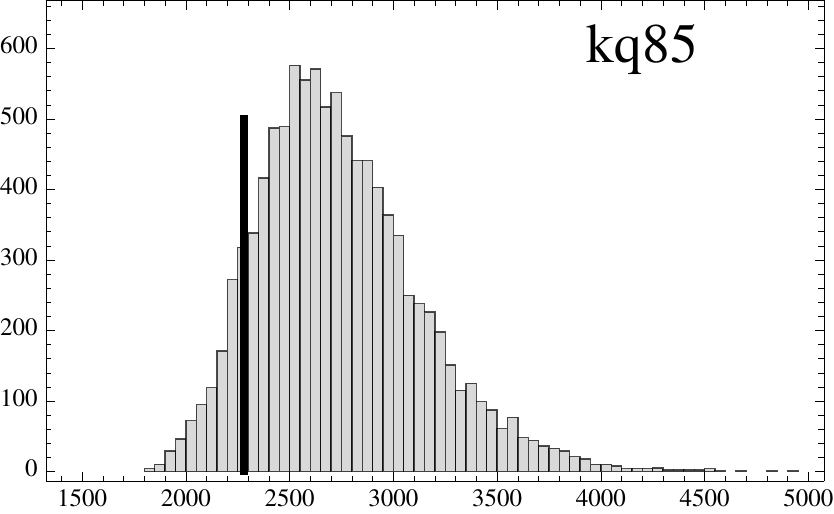}
\includegraphics[width=45mm]{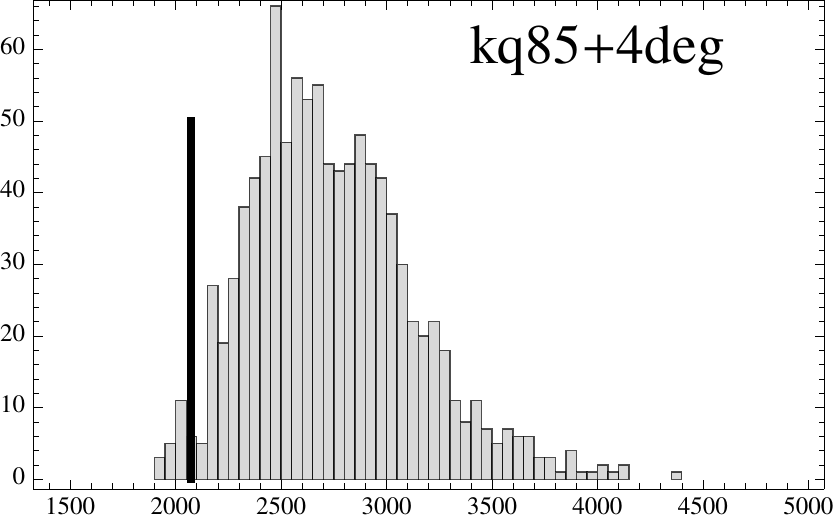}
\includegraphics[width=45mm]{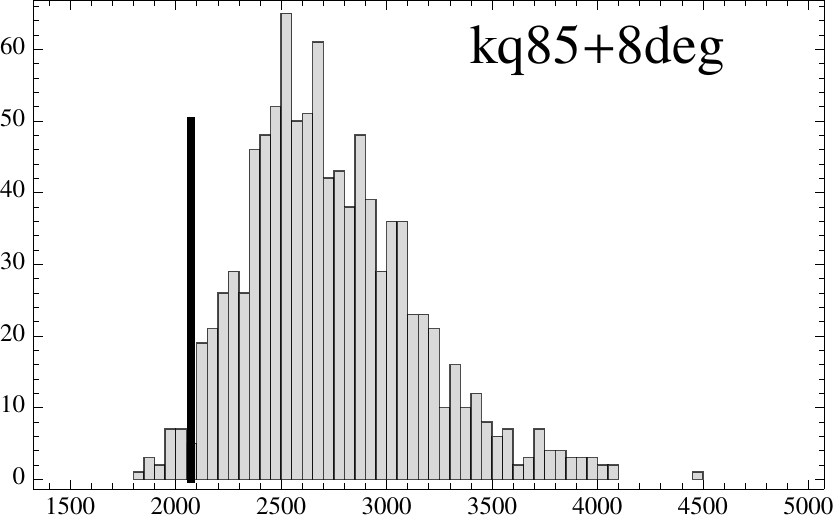}
\includegraphics[width=45mm]{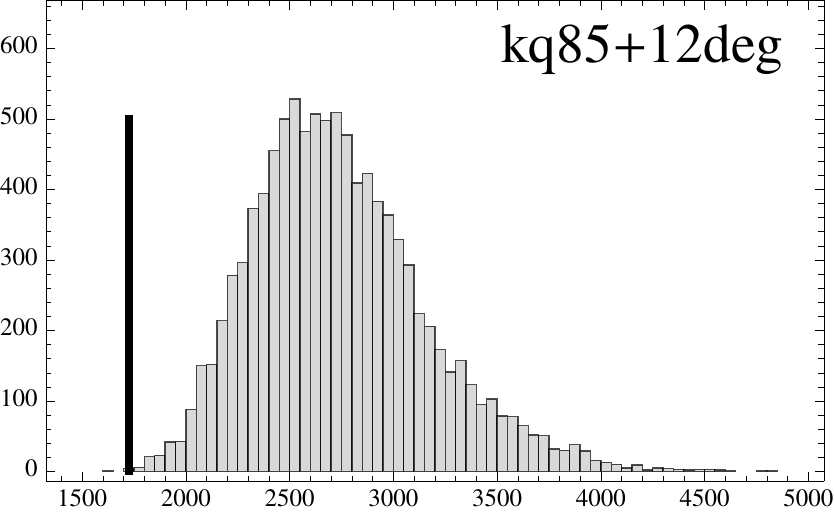}
\includegraphics[width=45mm]{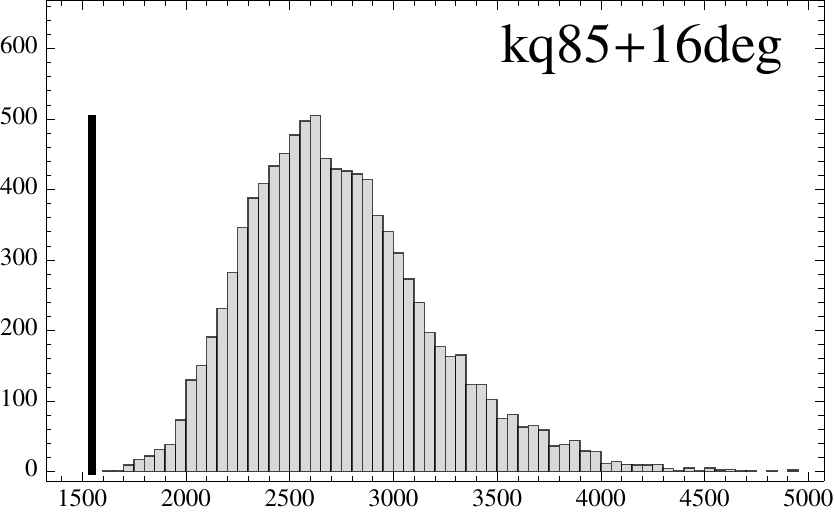}
\includegraphics[width=45mm]{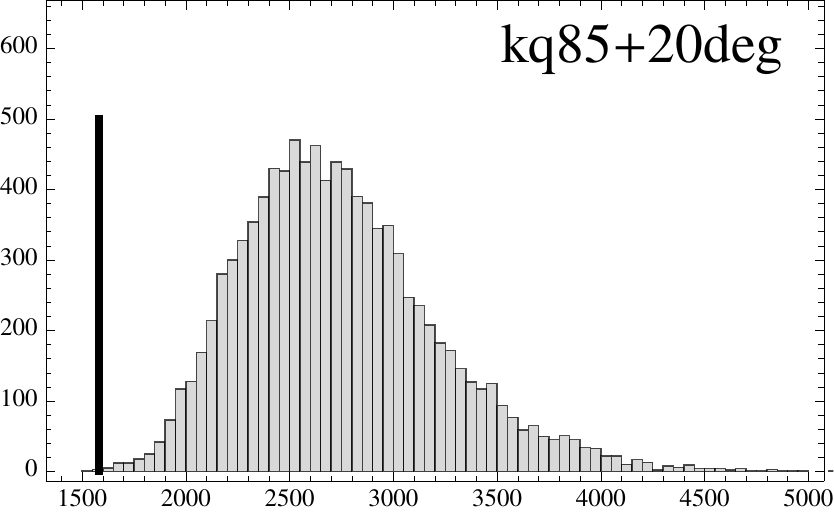}
\includegraphics[width=45mm]{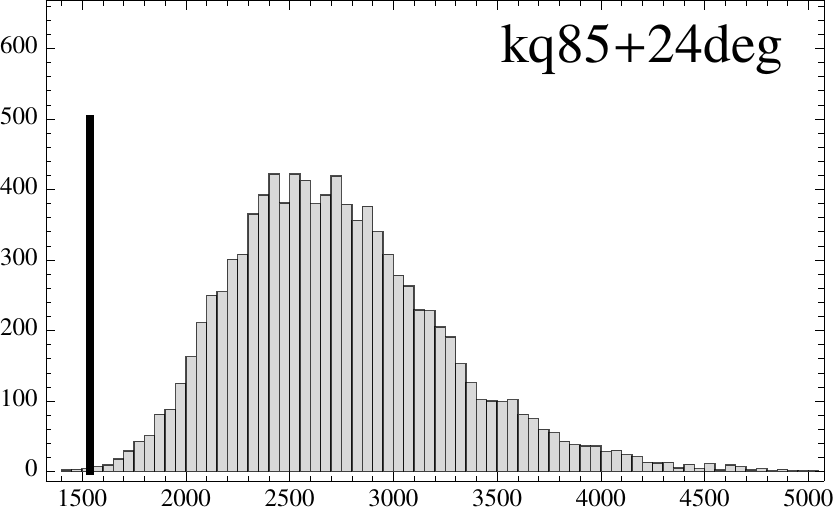}
\caption{TT Variance computed with Eq.~(\ref{VARTT}) with $\ell_{max}=32$ for the cases ``a'', ``b'', ``c'', ``d'', ``e'', ``f'' and ``g'' from top to bottom and left to right respectively. See also Table \ref{maskstabel}.
Each panel shows counts versus $\langle \delta T ^2 \rangle$. The vertical bars correspond to WMAP 9 year data.
Histograms for ``b'' and ``c'' cases are built with $10^3$ random extractions whereas all the other cases are built with $10^4$ random extractions.}
\label{sei}
\end{figure}

\begin{figure}
\centering
\includegraphics[width=60mm]{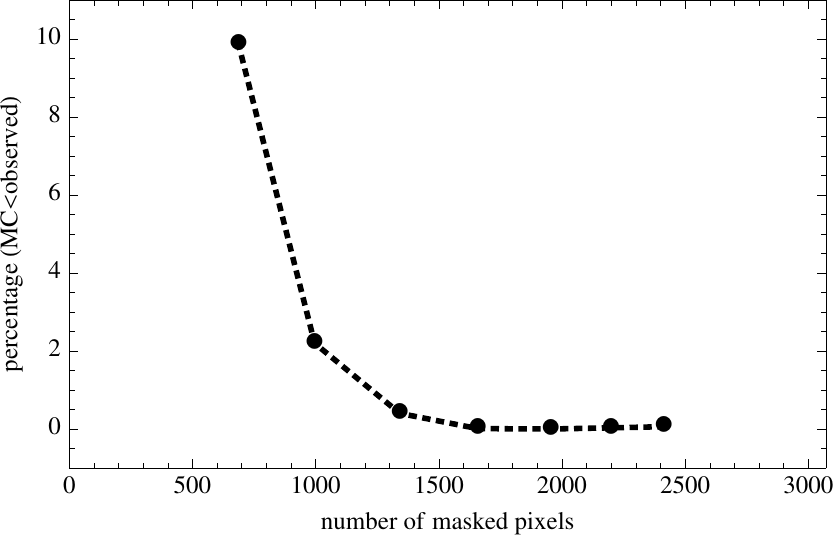}
\caption{Percentage of anomaly (i.e. lower tail probability) versus number of masked pixels.}
\label{seibis}
\end{figure}

In Fig.~\ref{sei} we show the variance $\langle \delta T ^2 \rangle$ computed through Eq.~(\ref{VARTT}) with $\ell_{max}=32$
for all the cases of Table \ref{maskstabel}. 
Each panel of Fig.~\ref{sei} shows the histogram obtained with MC simulations and the vertical bar corresponds to the WMAP data. 
Fig.~\ref{sei} shows that increasing the Galactic mask, the WMAP TT variance is more and more anomalous. The probabilities to obtain a smaller
value than WMAP from random simulations are $9.87\%$, $2.2\%$, $0.4\%$, $0.01\%$, $< 0.01\%$, $0.03\%$ and $0.06\%$  for the case ``a'', ``b'', ``c'', ``d'', ``e'', ``f'' and ``g'' respectively 
\footnote{All cases are analyzed with $10^4$ random simulations except for ``b'' and ``c'' that are studied with $10^3$ random simulations.}.
These percentages are plotted in Fig.~\ref{seibis} versus the number of masked pixels. 

\begin{figure}
\centering
\includegraphics[width=85mm]{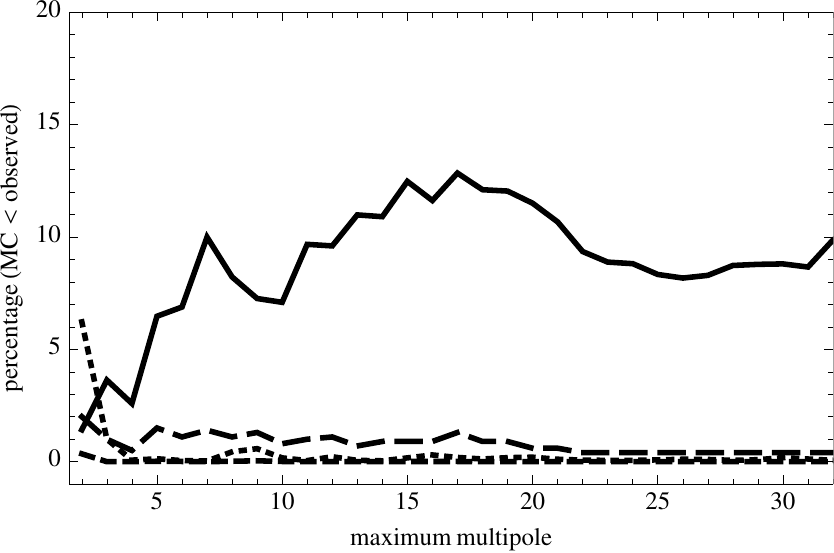}
\caption{The probability for TT variance to obtain a smaller value than the WMAP one for the cases ``a'' (solid line), ``c'' (long dashed line), ``e'' (dashed line) and ``g'' (dotted line) of Table \ref{maskstabel} as a function of $\ell_{max}$ with $\ell_{min}=2$.}
\label{sette}
\end{figure}

In Fig.~\ref{sette} we plot such a percentage for the cases ``a'', ``c'', ``e'' and ``g''  of Table \ref{maskstabel} as a function of $\ell_{max}$.
This Figure makes clear that the behavior given in Fig.~\ref{sei} for $\ell_{max}=32$ is in fact a general one
and it holds for every choice of $\ell_{max}$ from $2$ to $32$.
In particular, for the extension of $16$ degrees, i.e. case ``e'', the temperature variance is not consistent with $\Lambda$CDM model at less than $99.99\%$ C.L. in the
$\ell_{max}$ range $[25,35]$.

\begin{figure}
\centering
\includegraphics[width=85mm]{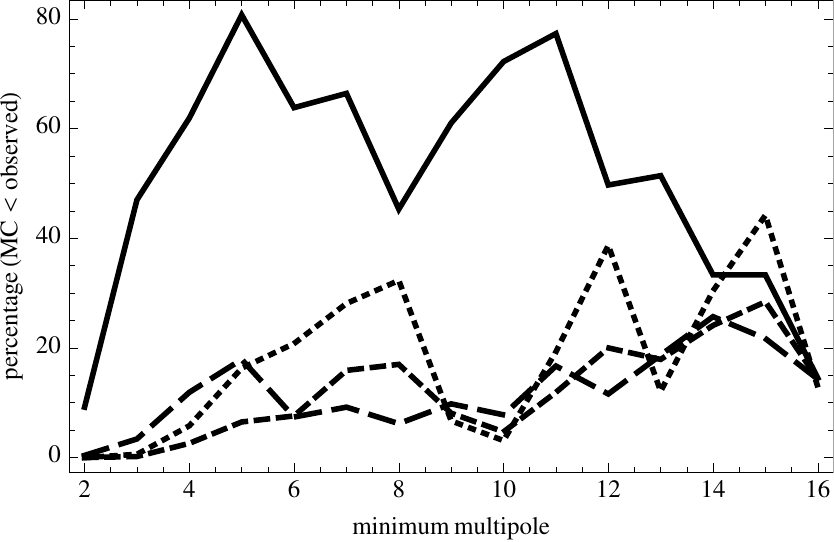}
\caption{The probability for TT variance to obtain a smaller value than the WMAP one for the cases ``a'' (solid line), ``c'' (long dashed line), ``e'' (dashed line) and ``g'' (dotted line) of Table \ref{maskstabel} as a function of $\ell_{min}$ with $\ell_{max}=32$.}
\label{otto}
\end{figure}

The low amplitude of the TT variance that we measure for the WMAP 9 year data, is dominated by the lowest multipoles.
In order to show this, we plot in Fig.~\ref{otto} the probability for the TT variance to obtain a smaller value than the WMAP one for the cases  ``a'' (solid line), ``c'' (long dashed line), ``e'' (dashed line) and ``g'' (dotted line) of Table \ref{maskstabel} as a function of $\ell_{min}$ with $\ell_{max}=32$. Fig.~\ref{otto} shows that excluding the quadrupole and the octupole the anomaly disappears.

\subsection{Impact of the smoothing procedure in the temperature analysis}
\label{impact}

\begin{figure}
\centering
\includegraphics[width=60mm]{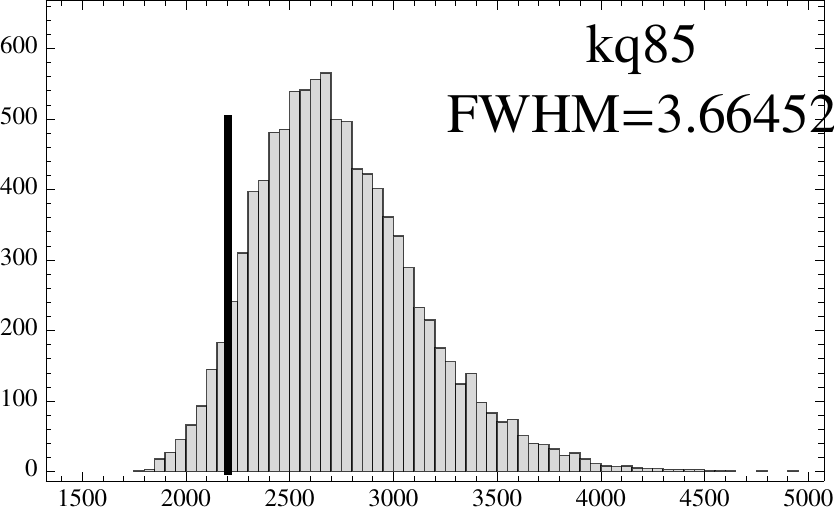}
\includegraphics[width=60mm]{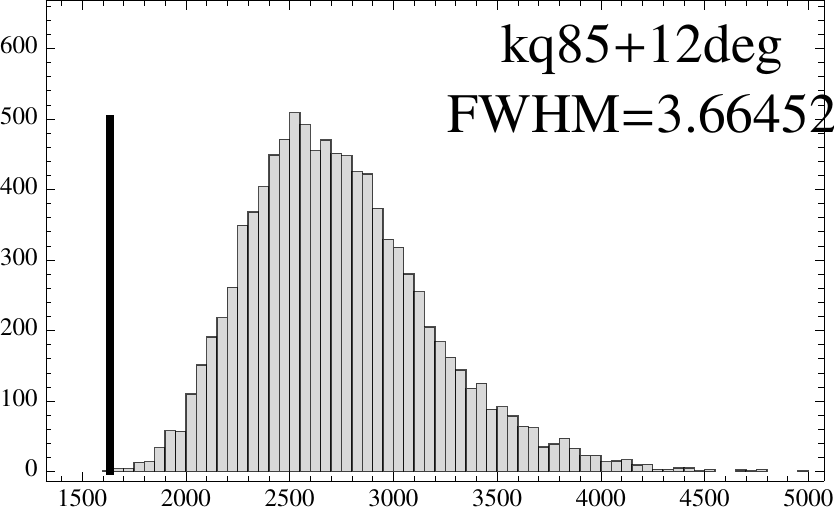}
\caption{TT Variance computed with Eq.~(\ref{VARTT}) with $\ell_{max}=32$ for the cases ``a'' and ``d'' from left to right respectively. See also Table \ref{maskstabel}.
Each panel shows counts versus $\langle \delta T ^2 \rangle$. The vertical bars correspond to WMAP 9 year data of the ILC map
pixellized at $N_{side}=16$ with a smoothing corresponding to the scale of the pixel at $N_{side}=16$. These panels are built with $10^4$ random extractions.}
\label{figuraimpact}
\end{figure}

We check here the stability of our findings against the degradation procedure of the Temperature map, needed to perform the analysis 
at low resolution. Specifically we have degraded the WMAP 9 year ILC map to $N_{side}=16$ performing a smoothing with $FWHM=3.66453 ^{\circ}$
which corresponds to the angular scale of a pixel at $N_{side}=16$.
Thanks to $10^4$ MC simulations we have re-computed Eq.~(\ref{VARTT}) for cases ``a'' and ``d'' with $\ell_{max}=32$, see Fig.~\ref{figuraimpact}.
The probability to obtain a smaller value than WMAP from the random simulations is $5.96\%$ and $0.01\%$, for the case ``a'' and ``d'' respectively.
Our results are proven to be stable.

\subsection{Polarization analysis}
\label{polarizationanalysissection}

\begin{figure}
\centering
\includegraphics[width=85mm]{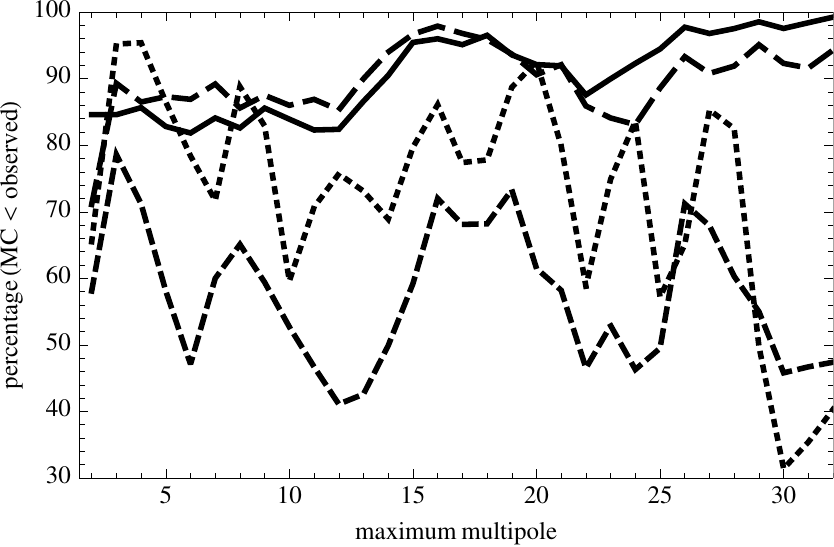}
\caption{The probability for  $\langle Q ^2 \rangle$ to obtain a smaller value than the WMAP one for the cases ``a'' (solid line), ``c'' (long dashed line), ``e'' (dashed line) and ``g'' (dotted line) of Table \ref{maskstabel} as a function of $\ell_{max}$.}
\label{nove}
\end{figure}

We have repeated for polarization the same analysis we have performed for temperature, see Eqs.~(\ref{VARQQ}),(\ref{VARUU}), finding no evident anomaly.
For sake of brevity we report in Fig.~\ref{nove} the analogous plot for polarization, already given in Fig.~\ref{sette} for temperature.
Even if in a regime that is noise dominated already at $\ell \gtrsim 10$, it is interesting to notice that increasing the polarization mask, the percentages move towards less anomalous values in
(almost) a monotonic way with respect to the sky fraction.

\section{Discussions and Conclusions}
\label{conclusions}


Using a QML estimator, we have provided new APS estimates of WMAP 9 year data at large angular scales and studied their dependence on Galactic masking.
We analyzed all six CMB spectra. 

By using realistic MC simulations we have found that the lowest multipoles of the TT spectrum decrease in amplitude as the Galactic mask is increased. 
This indicates that the temperature anisotropies around the Galactic mask behave statistically differently from the anisotropies at high Galactic latitude.
For example, extending the kq85 mask by 12 degrees, which covers $\sim 54\%$ of the sky, the temperature variance is not consistent with $\Lambda$CDM model at $99.94\%-99.99\%$ C.L. in the
$\ell_{max}$ range $[32,40]$. It is interesting to note that the trend is clearly monotonic with the sky masking up to the case ``d'' where the effect nicely saturates, see Fig.~\ref{seibis}.

We have also checked that the degradation procedure to obtain the low resolution WMAP 9 year ILC map does not impact our findings. 
To address this issue, we have degraded the WMAP 9 year ILC map to low resolution (corresponding to $N_{side}=16$) performing a
different smoothing\footnote{We remind that our results presented in Subsection \ref{temperatureanalysissection} are based on the WMAP 9 year ILC map smoothed with FWHM=$9.1285$ degrees, see Section \ref{dataset}.}
and we have re-evaluated for this map, Eq.~(\ref{VARTT}) in cases ``a'' and ``d''. 
Our results are also stable against the smoothing procedure, and therefore there is no evident spurious effect
that is introduced in this data manipulation step.
Furthermore we have verified that the lowest multipoles are responsible of such a variance anomaly.

Though based on different estimators, our findings are in agreement with the analysis performed in \citep{Cruz:2010ud}, for WMAP 5 and 7 data, and 
with \citep{Monteserin:2007fv}, where WMAP 3 year data is considered.
We do not find evidence of a similar effect in the other polarized spectra. This is not surprising since these are noise dominated already at $\ell \sim 10$.

Recently {\sc Planck} analysis of the variance for the temperature CMB anisotropies has appeared \citep{Ade:2013sta}.
We aim at analyzing these data as well along the same lines as given here for the temperature case since the pattern of TT APS at large angular scales is close to the WMAP 9 year data 
when the same mask is adopted \citep{Planck:2013kta}. Note that the lack of power at large angular scales claimed in \citep{Planck:2013kta} is not the same as the one discussed here, since
it involves different range of multipoles (see Fig.~39 of \citep{Planck:2013kta}).

If primordial, the low variance would constitute a challenge to the standard flat
$\Lambda$CDM cosmology emerging from a conventional inflationary expansion.
Among the many possible alternatives, which aim at explaining a low Sachs-Wolfe plateau,
an inflationary stage short enough to solve the big bang puzzles and also
disclose the preceding stage is an interesting possibility \cite{Contaldi:2003zv}, \citep{Destri:2009hn}, \citep{Dudas:2012vv}, \citep{Downes:2012gu} and \citep{Sagnotti:2013ica}.
Other possibilities are given by compact topology of the Universe: these models lead to similar effect in the Temperature maps but they also predict the existence of specific patterns like ``circles-in-the-sky'' that have not been detected, see e.g. \cite{Bielewicz:2011jz}.

We stress that in this paper we have tested a given model, i.e. the $\Lambda$CDM model, against WMAP 9 data in a pure frequentist fashion. However, as a proof of concept, we wish to provide a Bayesian analysis in the simplified case of Harrison-Zeldovich primordial spectrum, although disfavored by current observations, for which the variance of temperature fluctuations plays the role of a fundamental parameter (see Appendix \ref{HZ}). Our test shows that the posterior distribution for the variance moves towards smaller values, eventually saturating to a threshold value, for a more and more aggressive masking procedure. Therefore, we conclude that the bayesian and frequentist approach are qualitatively in very good agreement.



\acknowledgments A.G. wishes to thank Belen Barreiro, Marcos Cruz, Enrique Martinez-Gonzales and Patricio Vielva  for fruitful conversations.
We acknowledge the use of computing facilities at NERSC (USA) and CINECA (ITALY). 
We acknowledge use of the HEALPix (Gorski et al. 2005) software and analysis package for deriving the results in this paper. 
We acknowledge the use of the Legacy Archive for Microwave Background Data Analysis (LAMBDA). 
Support for LAMBDA is provided by the NASA Office of Space Science. 
Work supported by ASI through ASI/INAF Agreement I/072/09/0 for the Planck LFI Activity of Phase E2 and by MIUR through PRIN 2009 (grant n. 2009XZ54H2).

\appendix

\section{Bayesian approach within HZ models. A proof of concept.}
\label{HZ}

\begin{figure}
\centering
\includegraphics[width=85mm]{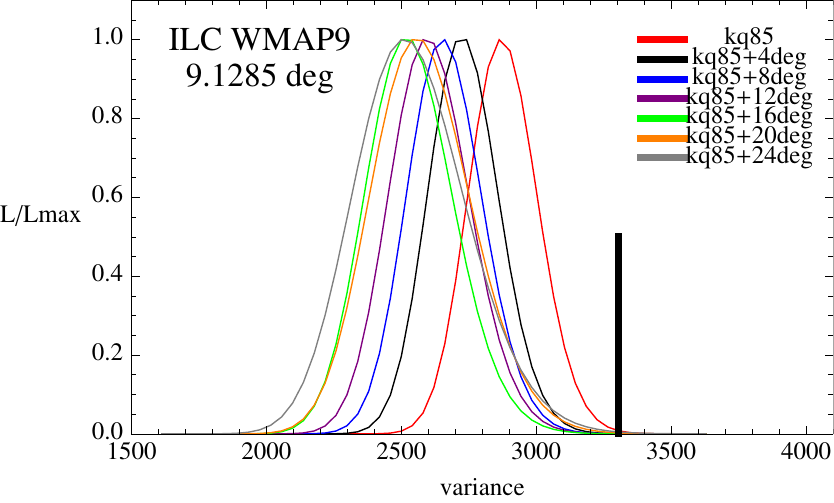}
\caption{Posterior distributions for the variance, $\sigma^2(\ell_{max}=32) \, \, [\mu K^2]$ obtained with various masks. }
\label{dieci}
\end{figure}

If the power spectrum of primordial perturbations is perfectly scale invariant (i.e. $n_{s}=1$, which defines the Harrison-Zeldovich model, henceforth HZ) then
the TT APS at low multipoles ($\ell \ll 90$) is flat and can be described by
\begin{equation}
{\ell (\ell +1) \over {2 \pi}} \, C_{\ell}^{TT} = A \, ,
\end{equation}
where $A$ is a constant and represents the amplitude of the primordial perturbations.
The HZ model is already ruled out by {\sc Planck} data \cite{Ade:2013uln} but is considered here just as a proof of concept.
Within the HZ model, the constant $A$ can be directly related to the variance, $\sigma^2(\ell_{max}) $, of the CMB anisotropies as
\begin{equation}
A = { 2 \, \sigma^2(\ell_{max}) \over {F(\ell_{max})}} \, ,
\end{equation}
where $F(\ell_{max}) = \sum_{\ell=2}^{\ell_{max}} {(2 \ell +1)}/ \ell (\ell +1)$.
In other words, for each given variance $\sigma^2(\ell_{max})$ it is possible to have a theoretical TT APS as
\begin{equation}
C_{\ell}^{TT} = { 4 \pi \sigma^2(\ell_{max}) \over { \ell (\ell +1) \, F(\ell_{max})}} = { 2 \pi \, A \over { \ell (\ell +1) }} \, ,
\label{SWmodel}
\end{equation}
and this greatly simplifies the Bayesian analysis.
Under the assumption of a flat prior on this model, one can sample on $\sigma^2(\ell_{max})$, in order to built a posterior ``slice'' on the variance, 
considering the likelihood expression
\begin{equation}
{\cal{L}} \propto \exp { (- x^T C^{-1} x \, / \,2 )} / \sqrt{ det (C) }
\, ,
\end{equation}
where $x$ is a vector containing the temperature CMB maps as observed by WMAP 9 year data and $C$
is the covariance matrix in pixel space defined as the sum $C = S(C_{\ell}) + N$ with
\begin{equation}
S(C_{\ell})_{ij} = \sum_{\ell} {{2 \ell +1}\over{4 \pi}} \, P_{\ell} (\hat i \cdot \hat j) \, C_{\ell}
\, ,
\label{signalcov}
\end{equation}
and $N$ being the noise covariance matrix 
\footnote{As a technical note, we wish to specify that  Eq.~(\ref{signalcov}) is always computed up to $\ell_{max}=64$ in order to build a well-defined matrix in pixels space.
This is done using Eq.~(\ref{SWmodel}). However the variance, over which we sample, is computed up to $\ell_{max}=32$ in order to make this case as close as we can with the frequentist analysis.}.
Sampling on $\sigma^2(\ell_{max}=32)$ for different masks we have obtained the posterior distributions shown in Figure \ref{dieci}.
In the same Figure, we also show a vertical bar which represents the variance for the SW model which is the closest to the $\Lambda$CDM model, defined through best fit of WMAP 9 data.  
In order to find such a model, here referred as $\Lambda SW$, we simply take the average of the band powers up to $\ell_{max}=64$ of the best fit of WMAP 9 data
finding
$
A^{\Lambda SW} \sim 1170 \, \mu K^2 \, 
$.
Plugging this number into $\sigma^2 = (A \, /2) \, F(\ell_{max}=32)$ we compute the expected value for the variance of this model, which reads
$\sigma^2_{\Lambda SW} \sim 3305 \, \mu K^2$.
Note that this value is larger with respect to the expected value of the variance in a pure $\Lambda CDM$ model which is $\sim 2737 \, \mu K^2$
for $\ell_{max}=32$.
See also the frequentist empirical distribution plotted in Fig.~\ref{sei}.

Figure \ref{dieci} shows that increasing the covering of the mask, the posterior for $\sigma^2$ moves towards smaller values. 
Therefore this Bayesian proof of concept shows a behavior qualitatively in agreement with the frequentist approach. 
This result strengthens the conclusions of the main text.
For completeness in Table \ref{SWtable} we show the percentage of compatibility of such a $\Lambda$SW model with respect to what obtained with WMAP 9 data.
This percentage is obtained with the following formula
\begin{equation}
\mbox{p} = 100 \, \int_{\sigma^2_{\Lambda SW}}^{\infty} p(\sigma^2) d \sigma^2
\, ,
\end{equation}
where $p(\sigma^2)$ is the normalized posterior distribution of $ \sigma^2$.
\begin{table}
\centering
\caption{Consistency with $\Lambda$SW model, see also the text. See also Table \ref{maskstabel}. }
\label{SWtable}
\begin{tabular}{cc}
\hline
Case & p (\%) \\
\hline
a & 0.109 \\
b & 0.005  \\
c & 0.003  \\
d & 0.003  \\
e &  0.003 \\
f & 0.035  \\
g & 0.077  \\
\hline
\end{tabular}
\end{table}
The values reported in Table \ref{SWtable} confirm the qualitatively trend of the frequentist approach.


\label{lastpage}

\end{document}